\documentclass[10pt,conference,a4paper]{IEEEtran}
\usepackage[english]{babel} 
\newcommand{\B}[1]{\boldsymbol{#1}}

\usepackage{amsmath}
\usepackage{amssymb}
\usepackage{graphicx}
\usepackage{xcolor}
\usepackage{epsfig}

\renewcommand{\d}[1]{\mathrm{d}#1}
\renewcommand{\r}[1]{\mathsf{#1}}
\newcommand{\E}[1]{\mathsf{E}\left[#1\right]}
\newcommand{\Eis}[1]{\mathsf{E}\left[\left|\r{i_s}(f)\right|^2\right] = }

\newcommand{\glna}{G_{\mathrm{LNA}}}

\newcommand{\cgd}{C_{\mathrm{gd}}}
\newcommand{\gs}{G_\mathrm{S}}
\newcommand{\gl}{G_\mathrm{L}}
\newcommand{\gm}{g_\mathrm{m}}
\newcommand{\go}{g_\mathrm{0}}
\newcommand{\gd}{g_\mathrm{d}}
\newcommand{\nf}{N_\mathrm{F}}
\newcommand{\no}{N_\mathrm{0}}
\newcommand{\phis}{\Phi_\mathrm{S}}
\newcommand{\phin}{\Phi_\mathrm{n}}
\renewcommand{\L}{\mathcal{L}}

\newcommand{\pdiff}[2]{\frac{\partial #1}{\partial #2}}

\graphicspath{{../thesis/images/matlab/}{images/}}

\begin{document}
\nocite{*}
\selectlanguage{english}
\title{Information Theoretic Analysis of Concurrent Information Transfer and Power Gain}
\author{{Fabian Steiner, Amine Mezghani and Josef A. Nossek}
\authorblockA{\\Institute for Circuit Theory and Signal Processing\\ Munich University of
Technology, 80290 Munich, Germany\\
E-Mail: \{fast, amme, jono\}@nws.ei.tum.de}\vspace{-0.8cm}}
\maketitle 
\begin{abstract}
In this paper, we analyze the fundamental trade-off between information transfer and power gain by means of an information-theoretic framework in communications circuits. This analysis is of interest as many of today's applications require that maximum information and maximum signal power are extracted (or transferred) through the circuit at the same time for further processing so that a compromise concerning the signal spectral shape as well as the matching network has to be found. To this end, the optimization framework is applied to a two-port circuit, which is used as an abstraction for a broadband amplifier. Thereby, we characterize the involved Pareto bound by considering different optimization problems. The first one aims at optimizing the input power spectral density (PSD) as well as the source and load admittances, whereas the second approach assumes the PSD to be fixed and uniformly distributed within a fixed bandwidth and optimizes the source and load admittances only. Moreover, we will show that additional matching networks may help to improve the trade-off.
\end{abstract}

\section{Introduction}
\label{sec:introduction}

The necessity of simultaneously achieving maximum information transfer (i.e. minimum noise figure) and maximum power gain occurs in communication circuits, such as the difficulty of achieving broadband matching for communication frontend circuits. In~\cite{grover}, the authors show that in the context of simultaneous power and information transfer achieving this goal always involves
finding a compromise concerning the spectral shape of the signal. Generally speaking, the maximum power transfer is obtained if all available power is concentrated on one single frequency, where the circuit presents the maximum power gain. In order to ensure a maximum information transfer (noise figure), the Waterfilling approach \cite{elements_it, gallager} shall be employed, requiring a certain amount of bandwidth that the spectrum is allowed to occupy. Additionally, the matching strategy of the circuit to the source and the load is usually  different with respect to the two aspects. 

Our work therefore aims at characterizing this trade-off, i.e., maximizing the amount of information being transfered through while simultaneously requiring a certain amount of signal power at the  output to obtain the Pareto-bound of this trade-off. A simple circuit model corresponding to such a situation will be presented in Section~\ref{sec:system_model}. It consists 
of a very basic version of a broadband amplifier which is an essential and common device in the receiver chain. The 
amplifier design problem also presents a trade-off between information transfer and maximum output power 
gain \cite{lna_design}. This issue is also known as the noise vs. the power matching design trade-off. In fact, 
circuit-designers aim at minimizing the SNR degradation caused by the braodband amplifier (measured by the noise figure), while 
assuring a certain output power gain such that the subsequent circuit components (mixers, ADCs) can work in an 
appropriate way. Due to the difficulty of defining the noise figure for broadband signals (frequency dependent), 
it is meaningful to utilize the Shannon's mutual information to measure the degree of degradation caused by the broadband amplifier, which leads 
to the proposed information theoretic framework.


In his very famous work~\cite{shannon}, Shannon was able to show that communication with a negligible error is possible as long as the communication
rate does not exceed the capacity of the channel. The channel capacity is hereby defined as the maximum of the mutual information over all possible input
stochastic processes: $C \triangleq \max_{p_\r{X}(x)}I(\r{X}, \r{Y})$. For practical applications an optimization of $I(\r{X}, \r{Y})$ given a constraint
concerning the maximum input power seems most reasonable. Solving this problem as stated in~\eqref{eq:waterfilling_prob} by a Lagrangian approach yields the
well-known water-filling structure of the PSD~\eqref{eq:waterfilling-sol}.

\begin{align}
& \max_{\phis(\omega)}
&& I(\r{X}, \r{Y}) = \int\limits_{-\infty}^\infty\frac{1}{2}\log_2\left(1+\frac{\phis(\omega)|H(\omega)|^2}{\Phi_n(\omega)}\right)\d{\omega} \nonumber \\
& \text{subject to}
&& \int\limits_{-\infty}^\infty \phis(\omega)\d{\omega} = P, \nonumber\\
&&& \phis(\omega) \geq 0, \quad\forall\omega\in\mathbb{R}.
\label{eq:waterfilling_prob}
\end{align}

\begin{align}
 \phis(\omega) = \left(\frac{1}{\lambda}-\frac{|H(\omega)|^2}{\Phi_n(\omega)}\right)^+.
 \label{eq:waterfilling-sol}
\end{align}

Hereby, $H(\omega)$ is the channel transfer function and $\Phi_s(\omega)$ and $\Phi_n(\omega)$ represent the signal and noise PSD respectively. The notation $(x)^+$ denotes the convenient way of expressing $\max\left(0, x\right)$. We will be able to discover strong similarities to our solution in the following sections, even if an additional power transfer constraint is present.

\section{System Model}
\label{sec:system_model}

\begin{figure}[ht]
 \begin{center}
 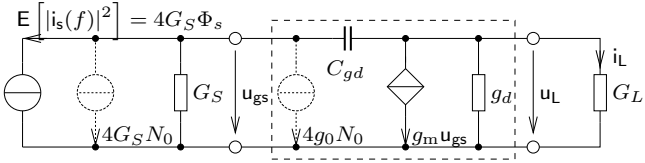
 \caption{Used small signal circuit model for the analysis.}
 \label{fig:circuit_model}
 \end{center}
\end{figure}

Fig.~\ref{fig:circuit_model} depicts the circuit model that is employed in order to obtain an insight into our 
ambition. In fact, it is a fairly simple small-signal model of a 2-port amplifier, but on the other hand it provides 
up to certain point an abstraction for a real broadband amplifier. Nevertheless, the analysis can be applied to other circuit models exhibiting such a trade-off. 

The transfer function $H(p) = \frac{\r{u_L}(p)}{\r{i_{s}}(p)/\gs}$ is in this case
\begin{align}
 H(p) = \frac{p\cgd\gs-\gm\gs}{\gs\gl+p\cgd(\gs+\gl+\gm)}.
 \label{eq:transfer_function}
\end{align}

With the help of~\eqref{eq:transfer_function}, the power gain can be obtained in a straightforward manner. The power gain -- in this context -- is defined as the ratio of the output power to the input power, which is frequency dependent and leads to the following form: 
 
{\footnotesize
\begin{align}
 \glna(\omega) &= \frac{\E{\left|\r{u_L}(\omega)\r{i_L}(\omega)\right|}}{\phis(\omega)} = \frac{\E{\left|\r{u_L}(\omega)\right|^2}\gl}{\frac{1}{4}\E{\left|\r{u_{s}(\omega)}\right|^2}\gs} = 4\frac{\gl}{\gs}\left|H(j\omega)\right| \nonumber \\
               &= \frac{4\gs\gl\left(\gm^2+\omega^2\cgd^2\right)}{(\gs\gl)^2 + \omega^2\cgd^2(\gs+\gl+\gm)^2}.
\label{eq:glna}
\end{align}}



Another influence which is of vital importance to our computations is the noise figure $N_F$ of the amplifier. This figure 
indicates the ratio of the SNR of the original input signal to the signal at the output port. Using the example of the circuit in Fig.~\ref{fig:circuit_model}, we obtain the noise figure when just considering the noisy input part of the circuit
\begin{align}
 \nf(\omega) = \frac{\mathrm{SNR}_\mathrm{in}(\omega)}{\mathrm{SNR}_\mathrm{out}(\omega)} = \frac{\frac{\gs\phis(\omega)}{\gs\no}}{\frac{\gs\phis(\omega)}{\gs\no +\go\no}} = 1 + \frac{\go}{\gs}.
 \label{eq:nf}
\end{align}
We note that, contrary to this example, the noise figure is usually frequency dependent. 
As we have seen before, the expression for the channel capacity depends on the term 
$\frac{\phis(\omega)|H(\omega)|^2}{\phin(\omega)}$, where $\phin(\omega)$ is the noise PSD seen at the output. 
 For the considered circuit, both the signal and the noise are subject to the same
transfer function. Therefore, the SNR at each frequency is given by

\begin{align}
 \frac{\phis(\omega)|H(\omega)|^2}{\phin(\omega)} = \frac{\phis(\omega)|H(\omega)|^2}{\nf\no|H(\omega)|^2} = \frac{\phis(\omega)}{\nf\no}.
\end{align}

\section{Mathematical Problem Formulation}
\label{sec:math_problem}

Having introduced all necessary prerequisites, we are now able to state the mathematical problem, which forms the 
foundation for our analysis. Before we tackle this issue in terms of precise mathematical
notation, the problem shall be revised in a brief summary: Our general goal is to maximize the information that can 
be transfered over a noisy channel. The Waterfilling algorithm provides the optimal way to distribute power between 
different independent channels given a maximum power constraint. However, as we have mentioned in the introduction, 
we would like to go one step beyond and see how the additional constraint of a desired minimum of power transfer 
affects the optimization results. 

To this end, the mathematical formulation would be
\begin{equation}
\begin{aligned}
& \max_{\phis(\omega), \gs, \gl}
&& \int\limits_{-\infty}^\infty \log_2\left(1+\frac{\phis(\omega)}{\nf\no}\right)\d{\omega} \\
& \text{subject to}
&& \int\limits_{-\infty}^\infty \phis(\omega)\d{\omega} = P, \\
&&& \int\limits_{-\infty}^\infty \phis(\omega)\glna(\omega)\d{\omega} \geq \eta P, \quad\eta \in\mathbb{R}^+,\\
&&& \phis(\omega) \geq 0, \quad\forall\omega\in\mathbb{R}.
\end{aligned}
\label{eq:optim_problem}
\end{equation}

As pointed out before, the objective function is the mutual information as introduced in (\ref{eq:waterfilling_prob}). The first constraint expresses that we do have a finite maximum available input power of $P$ that we are able to spread over the frequency depending on the form of $\phis(\omega)$. Now, the interesting part is concealed in the second constraint: We hereby enforce a certain amount of power to be transfered -- the exact
amount can be adjusted by setting $\eta$ correspondingly. A high value of $\eta$ indicates that we demand a large amount of power to be transfered. As a matter of fact, $\eta$ must be greater than one, in order to enable the amplifier to work as a device it was designed for.
We note that this optimization leads to the characterization of the trade-off (i.e. the Pareto bound) between information and power transfer of the two-port
circuit.

The maximization problem of~\eqref{eq:optim_problem} can be solved by means of the Lagrangian approach and the KKT conditions. The resulting
Lagrangian function reads as

{\footnotesize
\begin{multline}
 \L(\phis(\omega), \gs, \gl, \lambda, \mu, \kappa(\omega)) = -\int\limits_{-\infty}^\infty \log_2\left(1+\frac{\phis(\omega)}{\nf\no}\right)\d{\omega} +\\
 + \lambda \left( \int\limits_{-\infty}^\infty \phis(\omega)\d{\omega} - P\right)
 - \mu\left(\int\limits_{-\infty}^\infty \phis(\omega)\glna(\omega)\d{\omega} - \eta P\right)\\
 - \kappa(\omega)\phis(\omega)\notag.
\end{multline}
\label{eq:optim_phi_lagrange_function}}

Regarding the active set of inequality constraints one can conclude that the last two constraints of~\eqref{eq:optim_problem} have to be active (i.e. $\mu \neq 0$, $\kappa(\omega) \neq 0$, as any other theoretically combinations of the KKT multipliers would contradict the principles of primal feasibility and complementary slackness~\cite{cvx_optim}.

\section{Characterization of the Trade-Off}
\label{sec:trade_off}
\subsection{Optimization with Regard to the PSD and the Source and Load Admittances}
\label{sec:optim_phi_gs_gl}

The necessary optimality conditions provide a system of equations that has to be solved

{\footnotesize
\begin{align}
  \nabla \L(\phis(\omega), \gs, \gl, \lambda, \mu, \kappa(\omega)) = \begin{pmatrix} \pdiff{\L}{\phis(\omega)}, \pdiff{\L}{\gs}, \pdiff{\L}{\gl}, \pdiff{\L}{\lambda},  \pdiff{\L}{\mu}\end{pmatrix}^T = \B{0}.
\end{align}}

Caculating $\pdiff{\L}{\phis(\omega)}$ and setting it to zero, we get $\phis(\omega)$
\begin{align}
 \phis(\omega) = \left(\frac{1}{\ln(2)\left(\lambda-\mu\glna(\omega)\right)} - \nf\no\right)^+. 
 \label{eq:phi_optim_phis}
\end{align}

This expression can now be used in the other equations to derive the optimum values of $\lambda, \mu$ and $\gs, \gl$. Please note that the existence of a non-zero value of the
drain conductance $\gd$ is crucial for the existence of the trade-off. 

\begin{figure}[ht]
 \centering
 \includegraphics[scale=0.45]{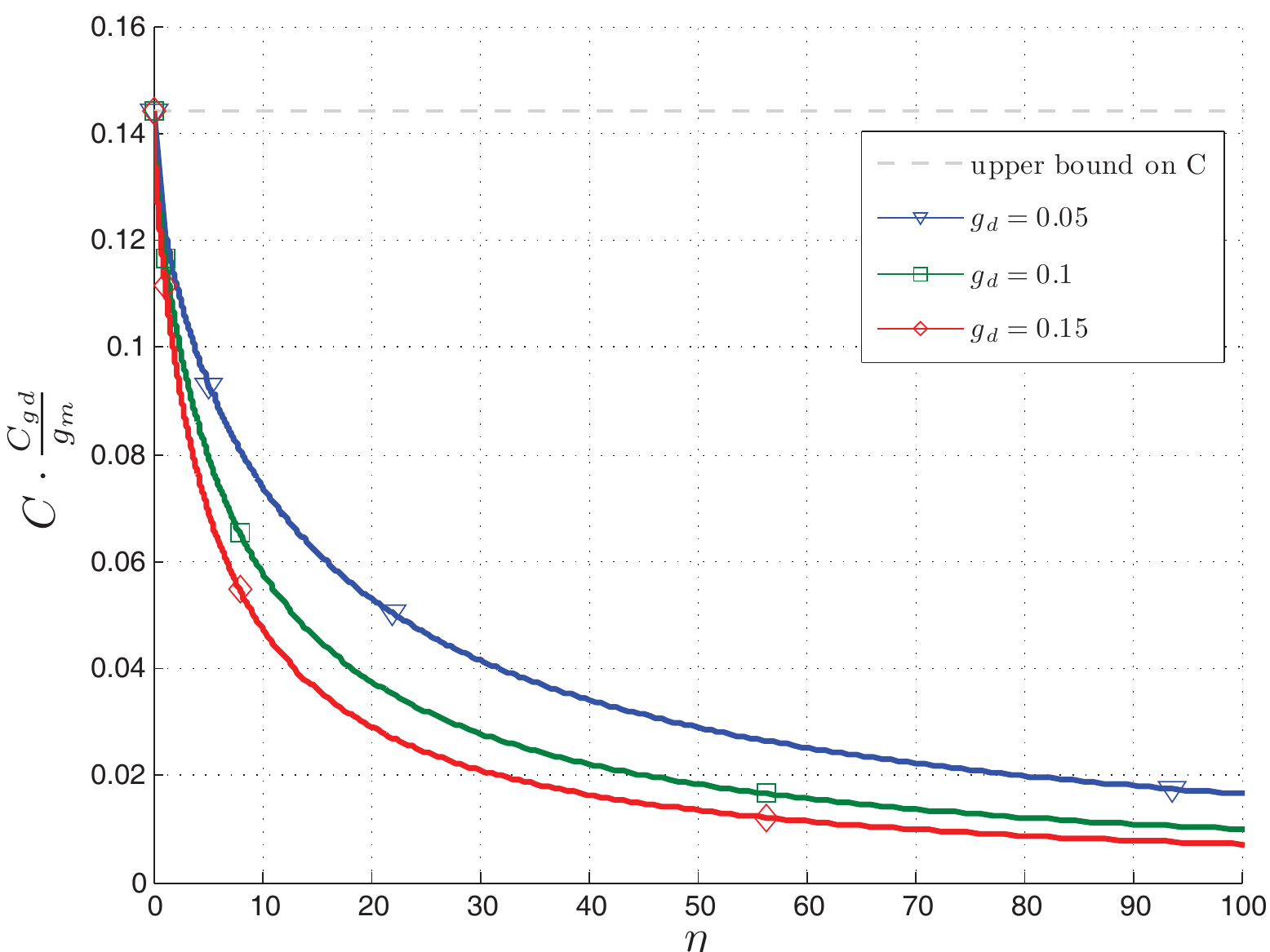}
 \caption{Plot of the power transfer vs. the channel capacity with different values of $\frac{g_{\mathrm{d}}}{g_{\mathrm{m}}} \in\{0.05,0.1,0.15\} $, $(P \frac{C_{\rm gd}}{N_0 g_{\rm m}}=0.1)$.}
 \label{fig:plot_eta_optim_phi_gs_gl}
\end{figure}

The $\eta-C$ curves in Fig.~\ref{fig:plot_eta_optim_phi_gs_gl} depict the trade-off between information and power transfer. Of course, one
recognizes that the curves are monotonously decreasing as the more power we demand to be transfered (larger $\eta$), the less information can be
transmitted. 


 
\begin{figure}[ht]
 \centering
 \includegraphics[scale=0.45]{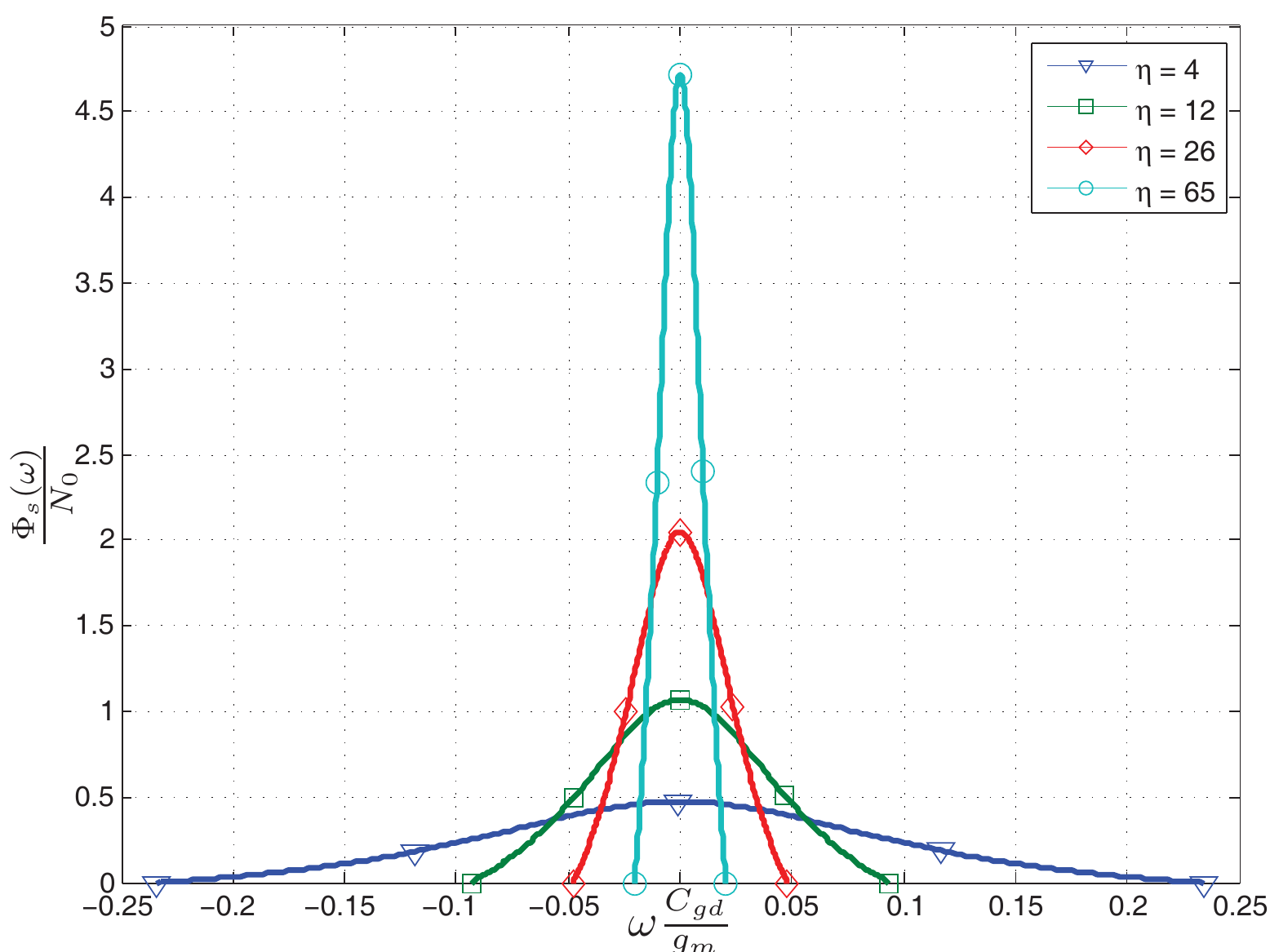}
 \caption{Resulting PSDs for different  power transfer factors $\eta$ ($\frac{g_{\mathrm{d}}}{g_m} =  P \frac{C_{\rm gd}}{N_0 g_{\rm m}}=0.1$).}
 \label{fig:plot_phi_optim_phi_gs_gl}
\end{figure}

To this end, Fig.~\ref{fig:plot_phi_optim_phi_gs_gl} provides an excellent insight into the trade-off between mutual information and power gain. It clearly depicts that the more power is demanded to be transfered (indicated by a larger value of $\eta$), the narrower (smaller bandwidth) and higher the graph becomes. If this process of increasing $\eta$ is continued, it is obvious that the resulting mathematical function converges to a Dirac distribution with zero bandwidth. In fact, this situation perfectly illustrates the problem of the bandwidth trade-off between both aspects as depicted in the introductory chapter.
\begin{figure}[ht]
 \centering
 \includegraphics[scale=0.45]{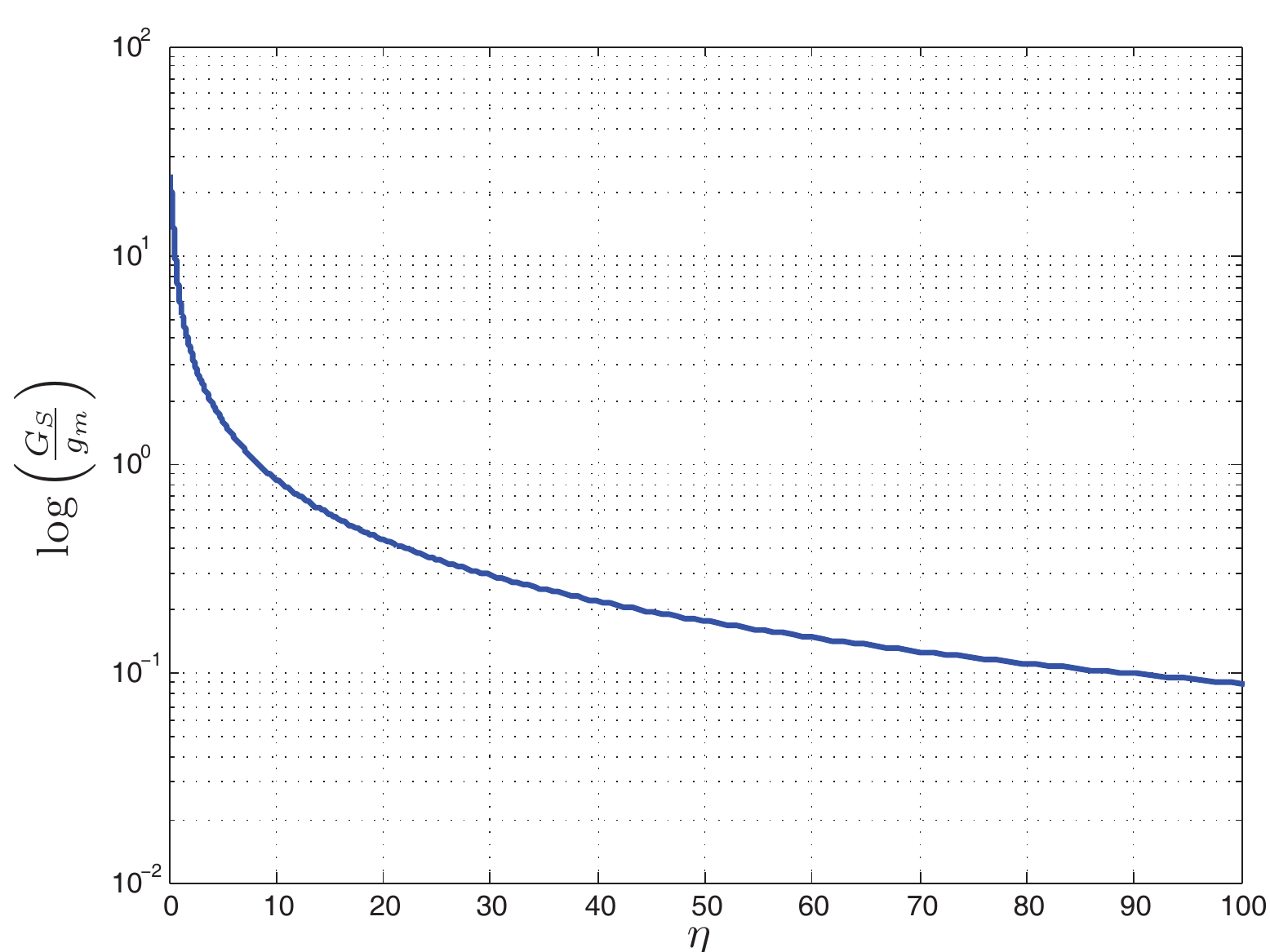}
 \caption{Plot of the resulting source admittance $\gs$ depending on the power transfer ($\frac{g_{\mathrm{d}}}{g_m} = P \frac{C_{\rm gd}}{N_0 g_{\rm m}}=0.1$).}
 \label{fig:plot_gs_optim_phi_gs_gl}
\end{figure}
\begin{figure}[ht]
 \centering
 \includegraphics[scale=0.45]{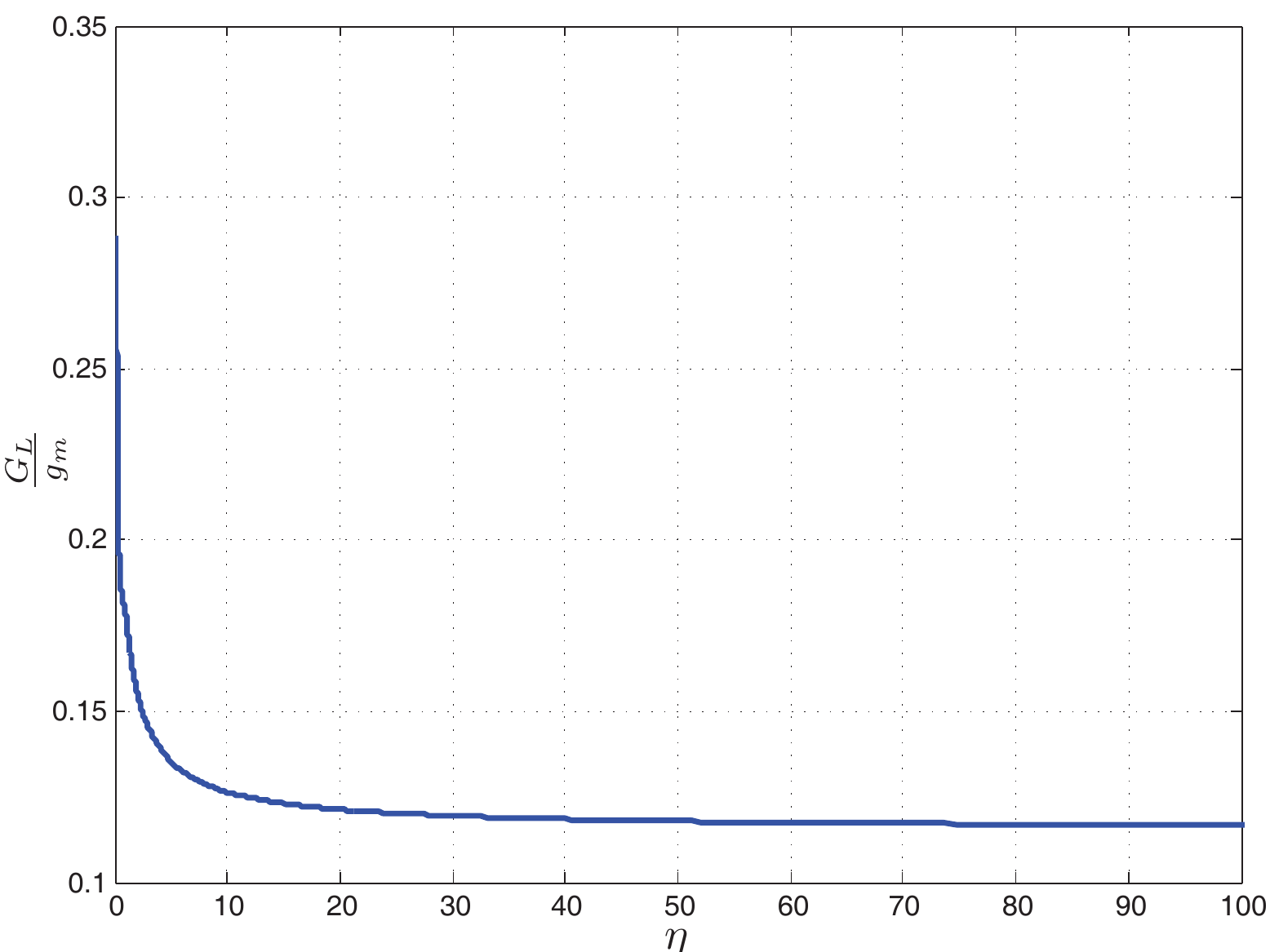}
 \caption{Plot of the resulting load admittance $\gl$ depending on the power transfer ($\frac{g_{\mathrm{d}}}{g_m} = P \frac{C_{\rm gd}}{N_0 g_{\rm m}}=0.1$).}
 \label{fig:plot_gl_optim_phi_gs_gl}
\end{figure}
\subsection{Optimization with Regard to Source and Load Admittances under a Uniformly Distributed PSD}
\label{sec:optim_phi_uniform_gs_gl}

\begin{figure}[ht]
 \centering
 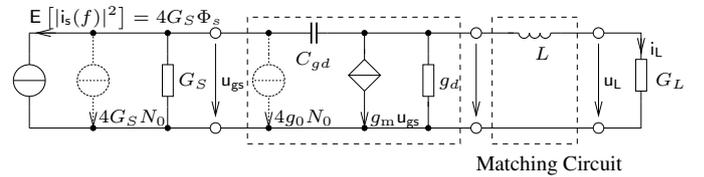
 \caption{Circuit model with additional matching component.}
 \label{fig:circuit_model_ext_matching}
\end{figure}

In contrast to our previous investigations, we now assume that the input power spectral density is fixed and uniformly distributed~\eqref{eq:def_phi_uniform}. This may apply
to reality when $\phis(\omega)$ is given by a certain application, which does not allow the spectral density to be modified. In such a scenario, the only free parameters are the source and load admittances which are therefore subject to on-going optimization. Besides, we also introduce an additional matching network at the output shown in Fig.~\ref{fig:circuit_model_ext_matching}.
\begin{align}
 \phis(\omega) = 
 \begin{cases}
  \frac{P}{\omega_B} & |\omega| \leq \frac{\omega_B}{2}\\
  0           & \text{otherwise}
 \end{cases}
 \label{eq:def_phi_uniform}
\end{align}

Consequently, the two necessary conditions with and without the matching network read respectively as
{\footnotesize
 \begin{align*}
  \nabla \L(\gs, \gl, \mu) = \begin{pmatrix} \pdiff{\L}{\gs}, \pdiff{\L}{\gl}, \pdiff{\L}{\mu}\end{pmatrix}^T = \B{0},
 \end{align*}}

and
{\footnotesize
 \begin{align*}
  \nabla \L(\gs, \gl, L, \mu) = \begin{pmatrix} \pdiff{\L}{\gs}, \pdiff{\L}{\gl}, \pdiff{\L}{L}, \pdiff{\L}{\mu}\end{pmatrix}^T = \B{0}.
 \end{align*}}

\begin{figure}[ht]
 \centering
 \includegraphics[scale=0.45]{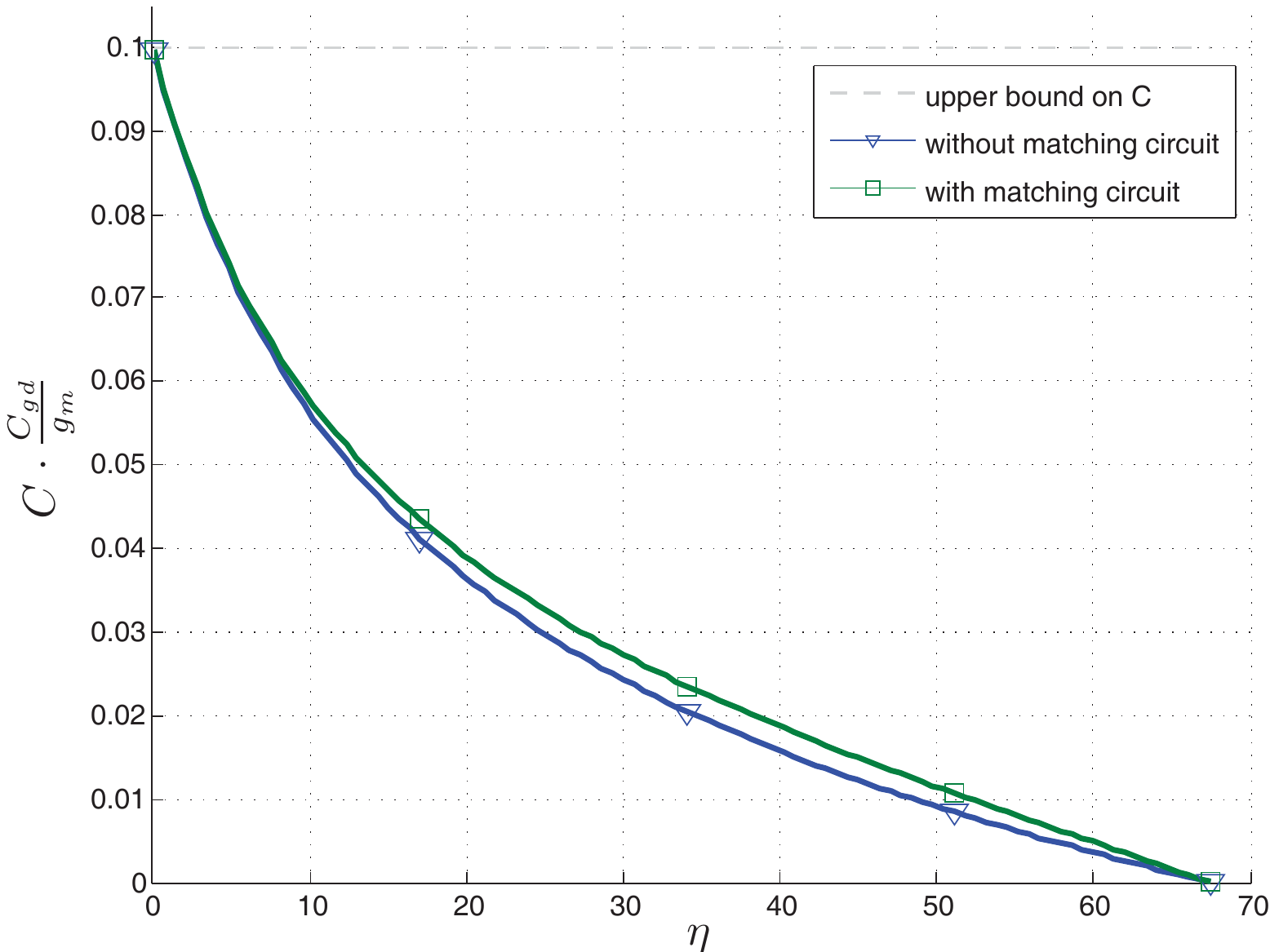}
 \caption{Plot of the power transfer vs. the channel capacity ($\frac{g_{\mathrm{d}}}{g_m} = \omega_B \frac{C_{\rm gd}}{g_{\rm m}} = P \frac{C_{\rm gd}}{N_0 g_{\rm m}}=0.1$).}
 \label{fig:plot_eta_optim_phi_uniform_matching}
\end{figure}

The numerical simulation results clearly prove the assumption that we had in mind when introducing an additional 
matching network. As Fig.~\ref{fig:plot_eta_optim_phi_uniform_matching} shows, the matching network is able to improve 
the trade-off by a certain extent. The extremal points (i.e. maximum channel 
capacity with zero power transfer and vice versa) obviously remain the same, but the 
overall Pareto bound becomes better due to the compensation of the gate-drain capacity.

\begin{figure}[ht]
 \centering
 \includegraphics[scale=0.45]{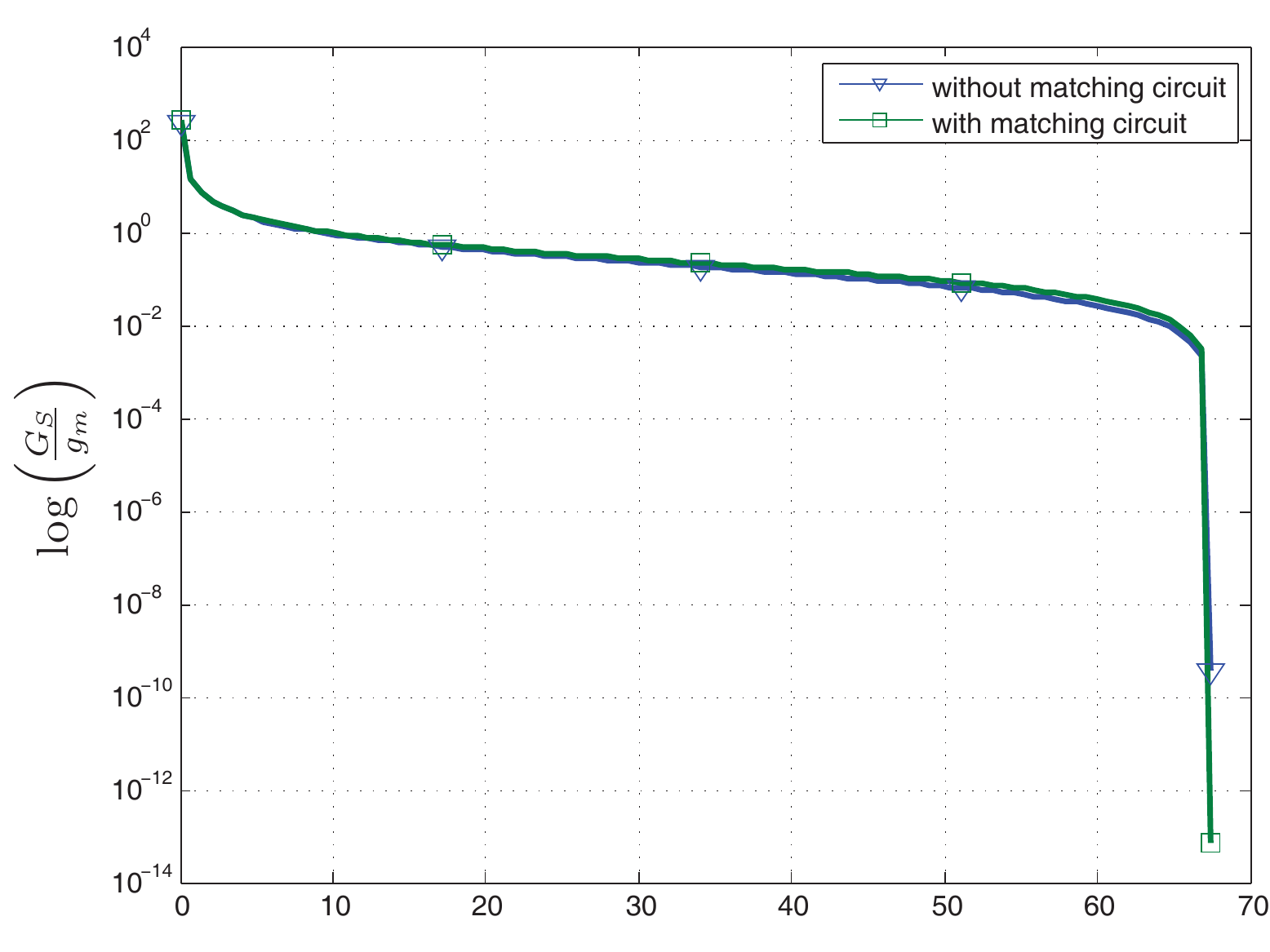}
 \caption{Plot of the resulting source $\gs$ depending on the power transfer ($\frac{g_{\mathrm{d}}}{g_m} = \omega_B \frac{C_{\rm gd}}{g_{\rm m}} = P \frac{C_{\rm gd}}{N_0 g_{\rm m}}=0.1$).}
 \label{fig:plot_gs_optim_phi_uniform_gs_gl_matching}
\end{figure}

The remaining figures can serve as a confirmation for the $\eta - C$ trade-off curve: 
Fig.~\ref{fig:plot_gs_optim_phi_uniform_gs_gl_matching} shows that the optimization yields a slightly larger value 
of $\gs$ which causes a higher capacity according to~\eqref{eq:nf}, \eqref{eq:optim_problem}.
The sudden decrease of the inductance value near the maximum gain can be explained if the 
derivative of the gain function with respect to $L$ is taken into account. As the source conductance approaches zero
(cf. Fig.~\ref{fig:plot_gs_optim_phi_uniform_gs_gl_matching}), the derivative begins to vanish, resulting in the fact that 
the influence of the inductance value is irrelevant.

\begin{figure}[ht]
 \centering
 \includegraphics[scale=0.45]{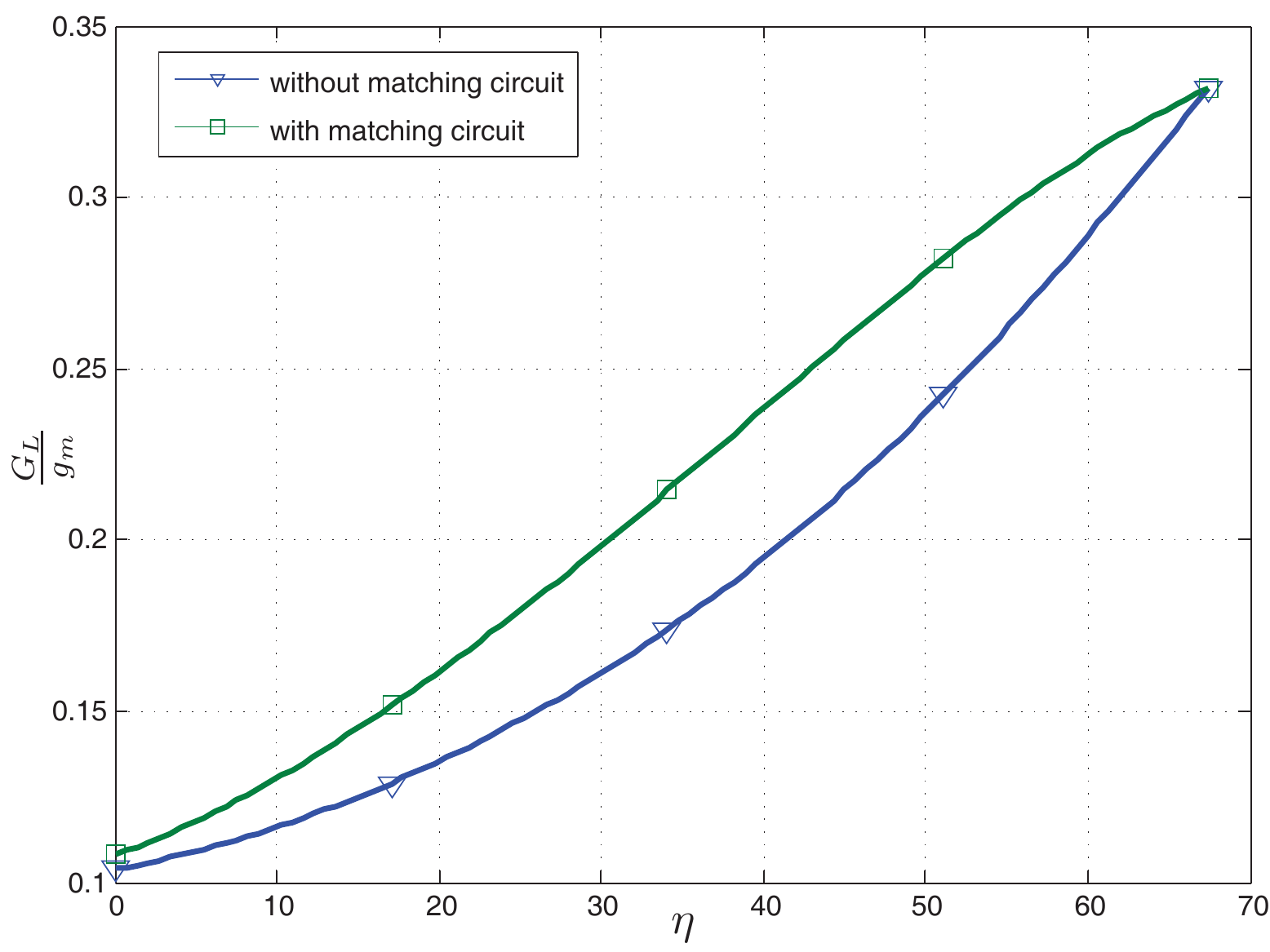}
 \caption{Plot of the resulting load admittance $\gl$ depending on the power transfer ($\frac{g_{\mathrm{d}}}{g_m} = \omega_B \frac{C_{\rm gd}}{g_{\rm m}} = P \frac{C_{\rm gd}}{N_0 g_{\rm m}}=0.1$).}
 \label{fig:plot_gl_optim_phi_uniform_gs_gl_matching}
\end{figure}

\begin{figure}[ht]
 \centering
 \includegraphics[scale=0.45]{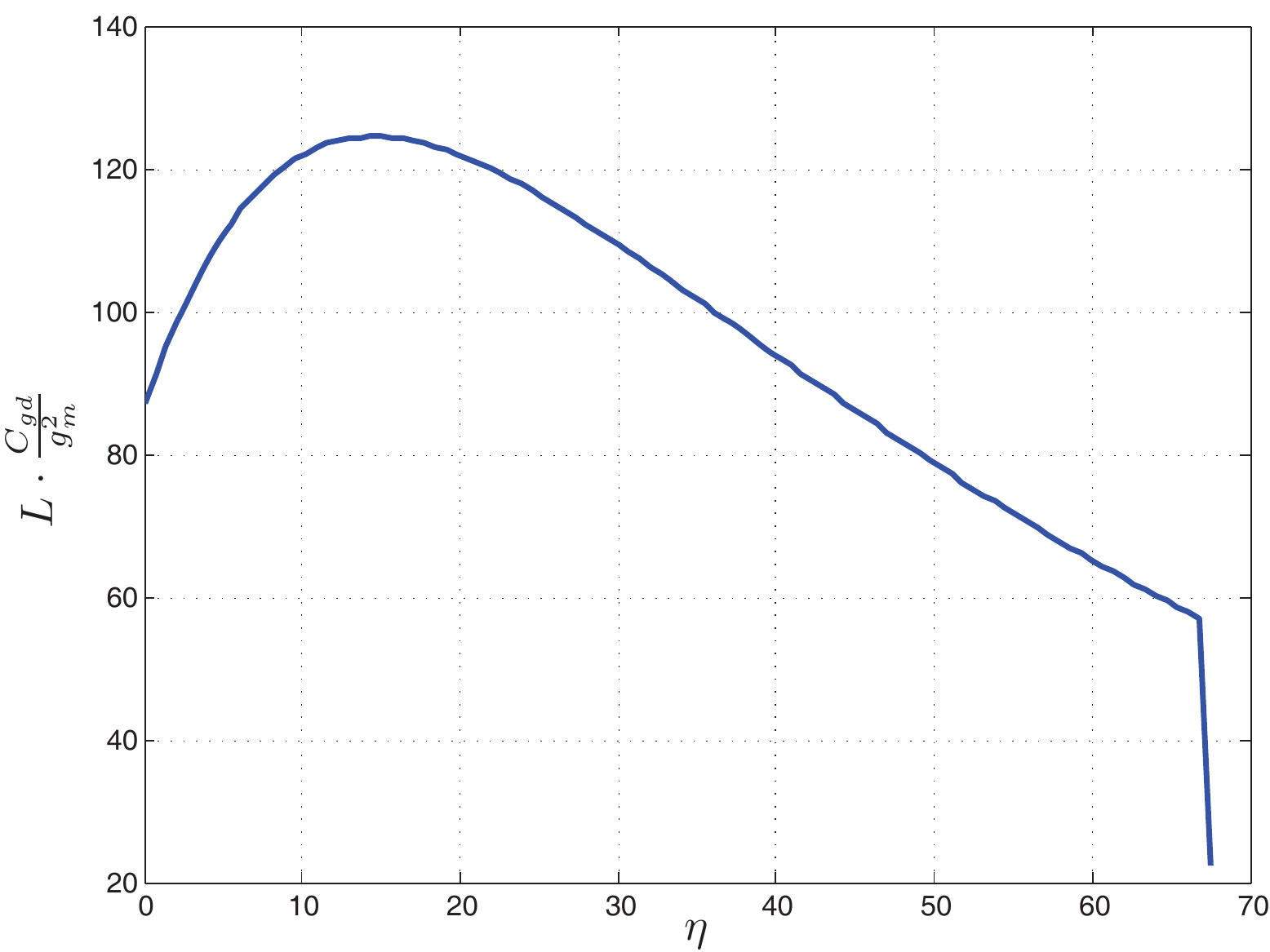}
 \caption{Plot of the resulting source matching inductance $L$ depending on the power transfer ($\frac{g_{\mathrm{d}}}{g_m} = \omega_B \frac{C_{\rm gd}}{g_{\rm m}} = P \frac{C_{\rm gd}}{N_0 g_{\rm m}}=0.1$).}
 \label{fig:plot_l_optim_phi_uniform_gs_gl_matching}
\end{figure}

\section{Conclusion}

We addressed the characterization of the trade-off between information and power transfer in designing communication circuits. As example, we considered a simple small signal equivalent circuit of a transconductance MOS broadband amplifier with internal noise source. Finding this trade-off consisted of maximizing the channel capacity given a constraint on the amount of power being transmitted at the same time. In the beginning, the optimization concentrated on optimizing the power spectral density as well as the source and load admittances. In the last section, we finally assumed the power spectral density to be uniformly distributed and optimized with respect to the source and load admittances. Besides, we found that an additional matching circuit may help to improve this trade-off significantly. Regarding future work, one should concentrate on further investigations involving matching circuits and considering more general circuits.

\bibliographystyle{IEEEbib}
\bibliography{iscas2012}

\end{document}